\title{A simple pendulum: Obtaining motion of pendulum bob from string tension time series}

\author{
Sparisoma Viridi\\
Nuclear Physics and Biophysics Research Division\\
Institut Teknologi Bandung, Bandung 40132, Indonesia\\
\and
Siti Nurul Khotimah\\
Nuclear Physics and Biophysics Research Division\\
Institut Teknologi Bandung, Bandung 40132, Indonesia\\
}

\date{\today}

\documentclass[12pt]{article}
\usepackage{graphicx}
\usepackage{amssymb}
\usepackage{wasysym}
\usepackage{gensymb}
\usepackage{hyperref}
\usepackage{breakurl}

\begin{document}
\maketitle

\begin{abstract}
Time series of string tension of a simple pendulum has not yet been a interesting motion information, even nowadays using a force tension sensor can be measured easily. A numerical procedure is presented how to obtain motion of pendulum bob from the string tension. Unfortunately, proposed procedure does not work. A quadratic term of $\omega$ loses the sign of $\omega$ that prevents the procedure to work. Applying sign of string tension into the procedure can produce the time series but only in positive value. The error for reproduced angular displacement is about 7.6 \%.
\end{abstract}

\section{Introduction}

Pendulum is an interesting system. It seems simple but it does not so. Even for a Galileo, it was crucial throughout his career \cite{Palmieri_2009}. The simple kind of pendulum is known as simple pendulum, which swings in small angle regime and has well known periode $T_0 = 2\pi\sqrt{l/g}$ \cite{Giancoli_2009, Benson_1996, Halliday_2006}. Complexity can be added to the system of a simple pendulum such as wide angle regime \cite{Belendez_2006}, elastic string \cite{Cuerno_1992}, or shaken pivot at where the string attached \cite{Schmitt_1998}. Especially for wide angle regime, Cristiaan Huygens in 1658 stated that a pendulum, which swings in very wide arcs of about 100 $\degree$, will introduce inaccuracy that is causing the swing period to vary with amplitude changes caused by small unavoidable variations in the driving or restoring force \cite{Edwardes_1970}. There are many attempts to formulate the periode of a simple pendulum for wide angle regime, for example, derived from energi consideration \cite{Lima_2006}, based on the arithmetic-geometric mean \cite{Carvalhaes_2008}, and by introducing a mutliplier \cite{Kidd_2002}. The simple experiment to measure the period automatically is by using a photogate sensor which registers only the time as the string is passing the sensor \cite{Khusnul_2011}. Information of string tension is not used to determine the motion but only to confirm the tension calculated from other motion parameters \cite{Unknown}. Based on experiment setup in \cite{Khotimah_2011} a calculation procedure to obtained motion of pendulum bob is reported in this work.

\section{Theory}

\begin{figure}[h]
\centering
\includegraphics[width=4cm]{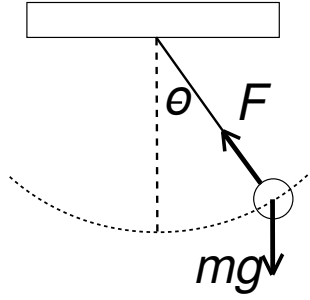}
\caption{\label{fg01} System of simple pendulum with all considered forces: tension force $F$ and earth gravitational force $mg$.}
\end{figure}

A simple pendulum system consists of a massless string with length $l$ and a pendulum bob with mass $m$ as illustrated in Figure \ref{fg01}. It is considered
in the system that air friction has no influence and the string can not be streched.

Instead of writing the equation of angular position of pendulum bob $\theta$

\begin{equation}
\label{eq01}
\frac{d^2\theta}{dt^2} + \frac{g}{l} \sin\theta = 0,
\end{equation}

which is derived from Newton's second law of motion in tangential direction, the law in radial direction is presented

\begin{equation}
\label{eq02}
\left(\frac{d\theta}{dt}\right)^2 + \frac{g}{l} \cos\theta - \frac{F}{ml} = 0,
\end{equation}

where string tension $F$ and angular position $\theta$ is function of time $t$. The simplest way to solved Equation (\ref{eq02}) to obtain solution of $\theta$ is by solving

\begin{equation}
\label{eq03}
\frac{d\theta}{dt} =  \pm \sqrt{\frac{F}{ml} - \frac{g}{l} \cos\theta}
\end{equation}

numerically with some initial conditions. Equation (\ref{eq03}) is a little bit different than that is proposed in \cite{Lima_2006} which is

\begin{equation}
\label{eq04}
\frac{d\theta}{dt} = \pm \sqrt{\frac{2g}{l} (\cos\theta - \cos\theta_0)},
\end{equation}

with initial conditions $\theta(0) = +\theta_0$ and $d\theta/dt(0)= 0$. Equation (\ref{eq03}) can be solved numerically. Using Euler method it can be written as

\begin{equation}
\label{eq05}
\theta_{\pm}(t + \Delta t) = \theta_{\pm}(t) \pm \sqrt{\frac{F(t)}{m l} - \frac{g}{l} \cos\theta_{\pm}(t)} \Delta t.
\end{equation}

Equation (\ref{eq05}), for both $\theta_{\pm}$, will be tested using artificial data produced by Equation (\ref{eq01}) dan (\ref{eq02}). Equation (\ref{eq01}) will give the following numerical equations to produce $\theta(t)$

\begin{eqnarray}
\label{eq06}
\alpha(t) = -\frac{g}{l} \sin\theta(t), \\
\label{eq07}
\omega(t + \Delta t) = \omega(t) + \alpha(t), \\
\label{eq08}
\theta(t + \Delta t) = \theta(t) + \omega(t),
\end{eqnarray}

with initial conditions $\theta(t) = \theta_0$ and $\omega = 0$ at $t = 0$. The produced time series of $\theta(t)$ will be used to produce $F(t)$ using

\begin{equation}
\label{eq09}
F(t) = m l \omega^2(t) + m g \cos \theta(t).
\end{equation}

\section{Results and discussion}

The $\theta(t)$ has been simulated with $l = 1 {\rm ~m}$, $g = 10 ~{\rm m/s^2}$, $\theta_0 = 0.5236 ~{\rm rad}$, $\omega_0 = 0$, $t_{\rm max} = 10 ~{\rm s}$, $\Delta t = 10^{-3} ~{\rm s}$, and $m = 0.2 ~{\rm kg}$. Using this time series the value of $\omega(t)$ is calculated using

\begin{equation}
\label{eq10}
\omega(t) = \frac{\theta(t) - \theta(t - \Delta t}{\Delta t}),
\end{equation}

and then the string tension $F(t)$ is calculated using Equation (\ref{eq09}), (\ref{eq10}), and the time series $\theta(t)$.

\begin{figure}[h]
\centering
\includegraphics[width=10cm]{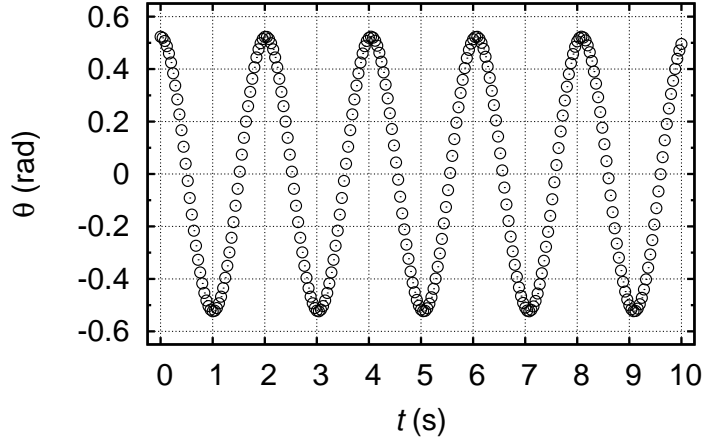}
\caption{\label{fg02} Produced times series $\theta(t)$ with $l = 1 {\rm ~m}$, $g = 10 ~{\rm m/s^2}$, $\theta_0 = 0.5236 ~{\rm rad}$, $\omega_0 = 0$, $t_{\rm max} = 10 ~{\rm s}$, and $\Delta t = 10^{-3} ~{\rm s}$.}
\end{figure}

\begin{figure}[h]
\centering
\includegraphics[width=10cm]{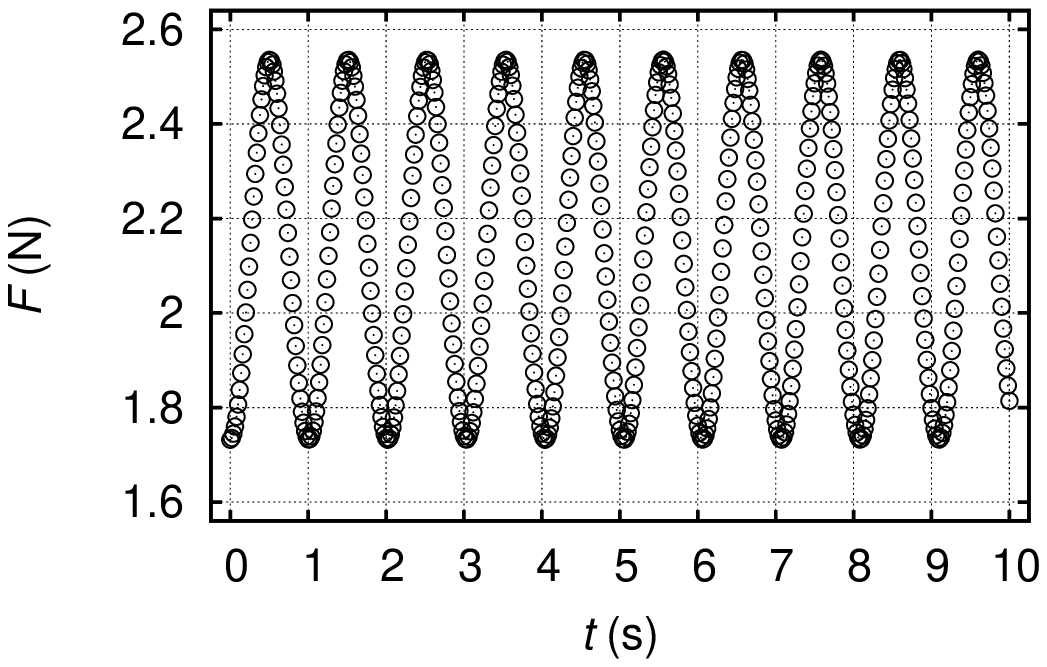}
\caption{\label{fg03} Produced times series $F(t)$ with $l = 1 {\rm ~m}$, $g = 10 ~{\rm m/s^2}$, $m = 0.2 ~{\rm kg}$ and $\Delta t = 10^{-3} ~{\rm s}$ from previous $\theta(t)$.}
\end{figure}

Figure \ref{fg02} shows the result of simulated $\theta(t)$ and the string force $F(t)$ produced from the $\theta(t)$ is given in Figure \ref{fg03}. Time series of $F(t)$ has periode double than the time series of $\theta(t)$. At time $t$ about 0.5 s the pendulum bob passes a point where it has minimum height. This point, also at $t$ about 1.5, 2.5, and other similar points, divide the range of one oscillation periode into two identical region where value of the string tension in these two regions with the same $|\theta|$ has the same value. That is way the periode of string tension $F(t)$ has periode double than the $\theta(t)$. In this model only magnitude of $F(t)$ is calculated that the direction always point to a point where the system is hung. The mathematical explanation is simpler, Equation (\ref{eq09}) gives always positive value from $\theta(t)$: the $\omega^2(t)$ and $\cos[\theta(t)]$.

\begin{figure}[h]
\centering
\includegraphics[width=10cm]{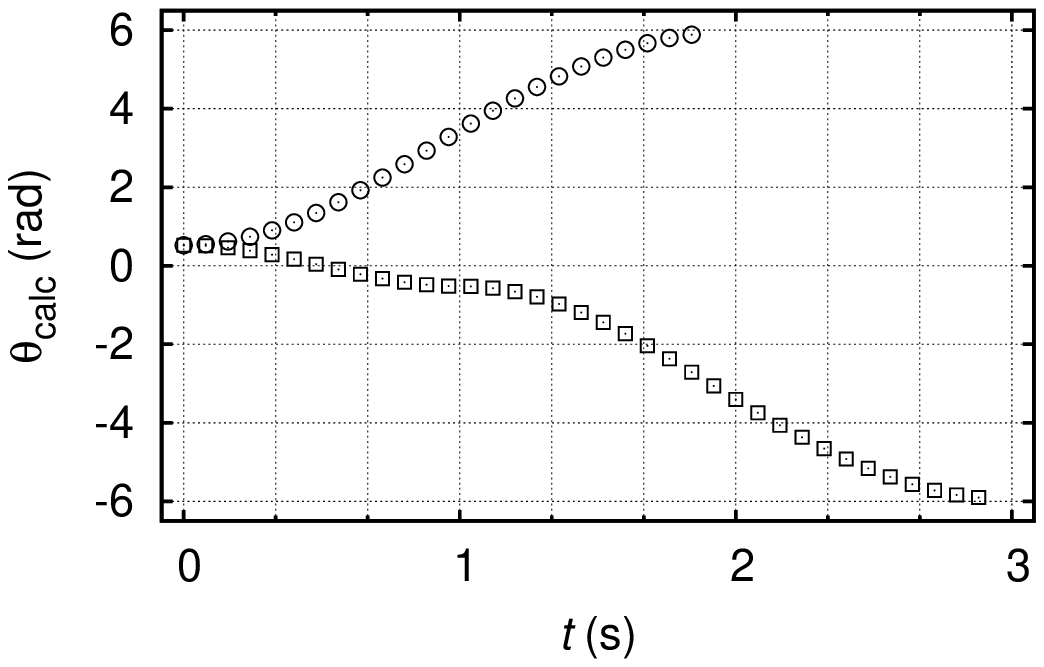}
\caption{\label{fg04} Produced times series $\theta(t)$ from $F(t)$ in Figure \ref{fg03} for: $\theta_{+}$ ($\circ$) and $\theta_{-}$ ($\Box$).}
\end{figure}

\begin{figure}[h]
\centering
\includegraphics[width=10cm]{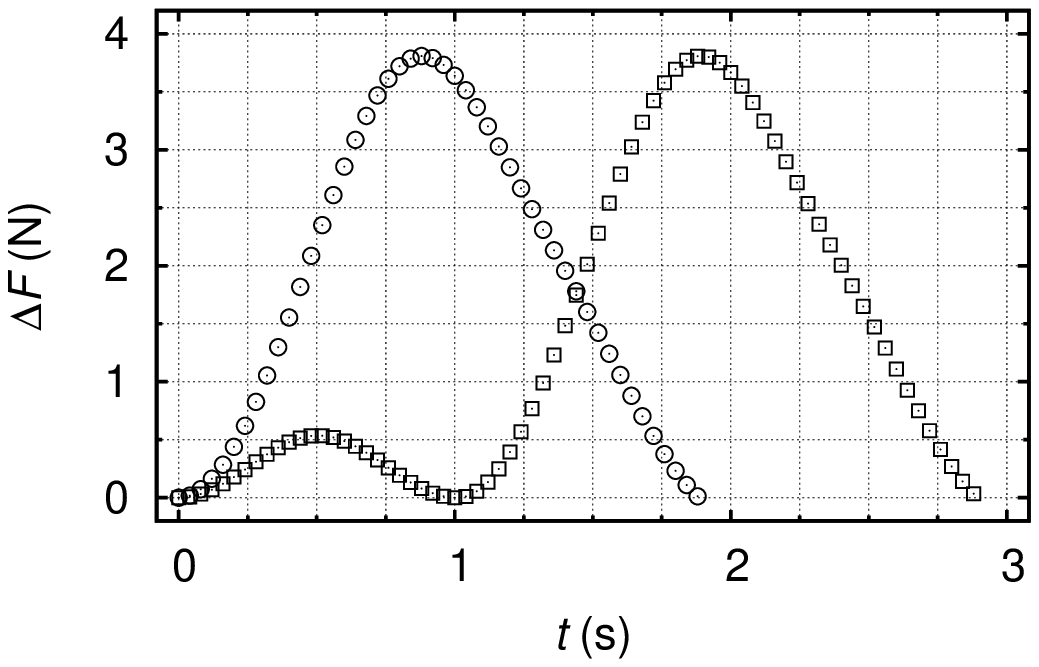}
\caption{\label{fg05} The difference between $F(t)$ and $F[\theta(t)]$ for: $\theta_{+}$ ($\circ$) and $\theta_{-}$ ($\Box$).}
\end{figure}

Unfortunately, direct application of Equation (\ref{eq05}) to calculate $\theta(t)$ from $F(t)$ give unintepreted results which has no physical meaning since value of $\theta(t)$ can be more than $\pm2 \pi$ rad. This result is shown in Figure \ref{fg04}. This could be lied on the $\pm$ sign in right side of Equation (\ref{eq05}). Further investigation shows that the calculated $\theta(t)$ does not produce the same $F(t)$ when it is used to calculate $F(t)$ using Equation (\ref{eq09}). It means that the integration used in this report does not guarantee that the results are self consistent. The different between $F(t)$ and $F[\theta(t)]$ which is represented as $\Delta F$ is given in Figure \ref{fg05}. A correction must be applied to overcome this problem.

An algorithm that could reduce the value of $\Delta F$ could probably give the right $\theta(t)$. This is the explanation why the previous result of $\theta(t)$ could have such a large value $\theta \in [-6:6]$ rad even the initial angle only 0.5236 rad. An iteratif self consisten field scheme is used to overcome this obstacle, but unfortunately can not handle this problem.

Other way is to modify Equation (\ref{eq05}) by notifying the behaviours of $F(t)$ and $\theta(t)$ from Figure \ref{fg02} and \ref{fg03}. Then Equation (\ref{eq05}) is modified into

\begin{equation}
\label{eq11}
\theta(t + \Delta t) = \theta(t) - \frac{|F(t + \Delta t) - F(t)|}{F(t + \Delta t) - F(t)} \sqrt{\frac{F(t)}{m l} - \frac{g}{l} \cos\theta(t)} \Delta t.
\end{equation}

\begin{figure}[h]
\centering
\includegraphics[width=10cm]{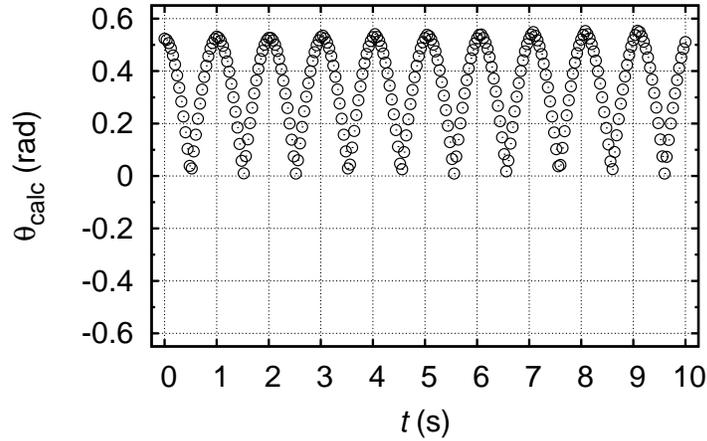}
\caption{\label{fg06} Produced $\theta_{\rm calc}(t)$ from $F(t)$  in Figure \ref{fg03}.}
\end{figure}

This modification can be implemented since the time series of $F(t)$ is already available. The result is given in Figure \ref{fg06}.

The difference between $\theta(t)$ and calculated $\theta_{\rm calc}(t)$, which is produced from $F(t)$, is defined as

\begin{equation}
\label{eq12}
\Delta \theta = |\theta(t)| - \theta_{\rm calc}(t).
\end{equation}

\begin{figure}[h]
\centering
\includegraphics[width=10cm]{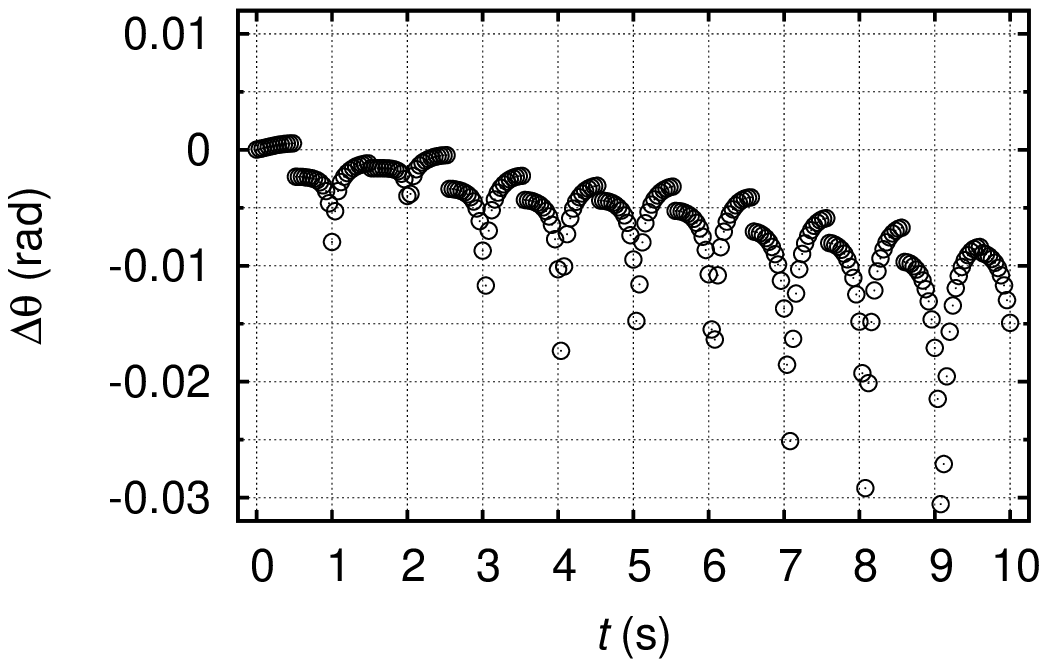}
\caption{\label{fg07} The difference between $\theta(t)$ and calculated $\theta_{\rm calc}(t)$.}
\end{figure}

Plot of $\Delta \theta$ against $t$ is given in Figure \ref{fg07}.

\section{Conclusion}

A numerical procedure to obtain time series of angular displacement $\theta(t)$ from string tension time series $F(t)$ of a simple pendulum has been reported. The similar Form of the equation that holds for tangential direction must be treated differently for radial direction since there is a quadratic term that hides the information of direction of angular velocity. A modification to the reqular form is shown and proved that it works. Only one minor problem remains, that the produced angular displacement has only positive values. The difference between orignal and reproduced angular displacement is below 0.04 rad for initial angular displacement 0.5236, or about 7.6 \% error.

\bigskip\bigskip
{\bf \Large Acknowledgements}\\ \\
Authors would like to thank to Research Division Research Grant from Institut Teknologi Bandung, Indonesia, in 2011 for partially support this work.

\bibliographystyle{unsrt}
\bibliography{manuscript}

\end{document}